\documentclass[twocolumn,showpacs,preprintnumbers,amsmath,amssymb,prl,epsf]{revtex4}
\usepackage{graphicx}

\begin{document}
\title{High-Visibility Multi-Photon Interference of Classical Light}
\author{I.~N.~Agafonov, M.~V.~Chekhova, T.~Sh.~Iskhakov, A.~N.~Penin}
\affiliation{Department of Physics, M.V.Lomonosov Moscow State
University,\\  Leninskie Gory, 119992 Moscow, Russia} \vskip 24pt

\begin{abstract}
\begin{center}\parbox{14.5cm}
{It is shown that the visibility of multi-photon interference for
classical sources grows rapidly with the order of interference. For
three-photon and four-photon interference of two coherent sources,
the visibility can be as high as $81.9$\% and $94.4$\%,
respectively, - much higher than the 'classical limit' of two-photon
interference (50\%). High-visibility three-photon and four-photon
interference has been observed in experiment, for both coherent and
pseudo-thermal light.}
\end{center}
\end{abstract}
\maketitle \narrowtext
\vspace{-10mm}

It is known that the nonclassical phenomena of two-photon
interference~\cite{Mandel1} and two- photon ghost diffraction and
imaging~\cite{Shih+Strekalov},~\cite{Shih+Pittman} have classical
counterparts. Two-photon interference of classical light has been
first discovered in the pioneering experiments by Hanbury Brown and
Twiss~\cite{HBT} and since then was observed with various sources,
including pseudothermal ones~\cite{Haner},
true thermal ones~\cite{Wu}, and coherent ones~\cite{Mandel2}.
Somewhat later, ghost imaging with classical light has been
demonstrated, both in the near-field and far-field
domains~\cite{Lugiato},~\cite{Boyd},~\cite{Shih}.
The only disadvantage of classical light with respect to two-photon
interference and two-photon ghost imaging, compared to two-photon
entangled sources, is the limited visibility, which is always below
$50$\%~\cite{Mandel3},~\cite{aspekty}. On the other hand, in
interference and imaging experiments with classical sources the
visibility is independent of the intensity, which can therefore be
arbitrarily high. Entangled sources, in contrast, should be
sufficiently weak to provide high-visibility interference: while the
visibility is close to 100\% for faint two-photon light, it drops
with the increase in the mean number of photons per
mode~\cite{NL},~\cite{Lugiato}.

In this paper we show that classical sources can provide much better
visibility if one passes to higher-order intensity correlations.
This fact, which has considerable importance for multi-photon
imaging and multi-photon lithography, is also remarkable in view of
the numerous recent experiments on higher-order correlations and
multi-photon interference~\cite{multiphoton}. Indeed, while in the
case of two-photon interference any experiment with the visibility
exceeding $50$\% can be interpreted as non-classical, no such
criterion is formulated for higher-order interference.  Our results
show that for multi-photon interference to be recognized as
nonclassical, its visibility should exceed really high values:
$81.9$\% in the three-photon case and $94.4$\% in the four-photon
one. These values are classical visibility limits for three- and
four-photon interference, respectively.

For our consideration, we chose the scheme of the two-slit Young's
interference experiment~\cite{generality}. This geometry was used in
many experiments on two-photon interference and two- photon ghost
imaging~\cite{Shih+Strekalov},~\cite{Shih+Pittman},~\cite{Haner},~\cite{Boyd},
~\cite{Lugiato},~\cite{Shih}. This time, however, we consider the
interference to be registered by three detectors instead of two
(Fig.1), each detector measuring the instantaneous intensity and the
triple photocount coincidences being counted by a coincidence
circuit. This is the standard experimental technique to measure the
third-order Glauber's intensity correlation function
(ICF)~\cite{GCF}. Let the two sources $A$, $B$ be classical ones,
having the same statistics, the same average intensities, and
independently fluctuating phases. The last condition provides the
'erasure' of the first-order interference in the far-field zone.
However, a simple calculation shows that the intensity correlation
functions will be sensitive to the positions of the detectors.
Passing to the normalized third-order correlation function for the
detectors placed at points 1, 2, 3,
\begin{equation}
g_{123}^{(3)}\equiv\frac{\langle I_1I_2I_3\rangle}{\langle
I_1\rangle\langle I_2\rangle\langle I_3\rangle}, \label{g3def}
\end{equation}
$I_j$ being the instantaneous intensity at point $j=1,2,3$, we
obtain that
\begin{equation}
g_{123}^{(3)}=\frac{g^{(3)}}{4}+
\frac{g^{(2)}}{2}[\frac{3}{2}+\hbox{cos}\phi_{12}+
\hbox{cos}\phi_{23}+\hbox{cos}\phi_{13}],
\label{g3}
\end{equation}
where $g^{(2)}, g^{(3)}$  are, respectively, the second-order and
third-order normalized ICFs for each of the two sources and
$\phi_{ij}\equiv\phi_{Ai}-\phi_{Aj}-\phi_{Bi}+\phi_{Bj}$,
$\phi_{Sd}$ being the phase gained by the radiation of source
$S=A,B$ on the way to detector $d=1,2,3$.

According to Eq.(\ref{g3}), the maximal visibility is provided by
the radiation sources with the minimal ratio of $g^{(3)}/ g^{(2)}$.
Among the classical sources, this 'visibility limit' is achieved for
coherent light and is equal to $81.9$\%. Thermal radiation gives a
lower visibility, $60$\%, which is still much higher than the
corresponding value in the case of two-photon interference ($33$\%).

In our experiment, the third-order Glauber's correlation function
was measured through the coincidence counting rate of three
detectors (Fig.2). As the radiation source, we used a frequency
doubled Q-switched YAG:Nd laser with the wavelength $532$ nm, pulse
duration $5$ ns, and the repetition rate $3$ kHz. Instead of two
slits, which should be very precisely matched in width to achieve
the maximum visibility of multi-photon interference, a single slit
of width $150 \mu$ was used, followed by a birefringent crystal
(calcite). The crystal split the beam into the ordinary one and the
extraordinary one; with the slit and the crystal placed between
crossed polarization (Glan) prisms, this configuration was
equivalent to two identical slits separated by a distance of $1.3$
mm. In the far-field zone, where the interference pattern was
formed, the radiation was attenuated using neutral-density filters
and fed into a three-arm Hanbury Brown-Twiss interferometer with
three photon-counting avalanche diodes and a triple-coincidence
circuit. Attenuation was necessary to keep the average number of
photocounts per pulse much less than one; otherwise, because of the
detectors dead-time effect, the photocount and coincidence rates
would be measured incorrectly. Gating of the registration electronic
system provided suppression of the dark noise by several orders of
magnitude. In order to scan the interference pattern, thick glass
plates were placed at the inputs of detectors 1 and 3. Spatial mode
selection was provided by $70 \mu$ apertures A1, A2, A3 placed in
front of each detector. By turning the glass plates, one could scan
the phase of either one or two detectors within the range $0\dots
6\pi$. The phase of detector 2 could be adjusted by moving the
aperture A2.

The varying relative phase of the two sources was introduced by
inserting an electro-optical modulator (EOM) after the calcite
crystal. The EOM was oriented in such a way that its bias voltage
induced a phase shift between the ordinary and extraordinary beams.
Applying an AC voltage with the amplitude $85$ V (slightly below the
quarter-wave one) and the frequency $50$ Hz, we erased the
interference pattern in the time-averaged intensity distribution.
Although this harmonic phase modulation was different from a
randomly varying relative phase of sources $A,B$ (Fig.1), it
resulted in the same time-averaged ICFs.

We studied third-order interference both for coherent and
pseudo-thermal radiation at the input. Pseudo-thermal light was
prepared by means of a rotating ground-glass disc placed after the
calcite crystal. The envelope of the third-order spatial ICF was
determined by the coherence radius of the radiation, which, in its
turn, depended on the sizes of the spots formed on the disc by the
ordinary and extraordinary beams, and its FWHM corresponded to
approximately two periods of the interference pattern. Because of
this, the visibility in the case of thermal light was considerably
less than the theoretical value.

The results for the coherent and pseudo-thermal cases are presented
in Fig.3 a,b, respectively. Experimental points correspond to the
normalized third-order Glauber's ICF; curves show the dependencies
given by Eq.(\ref{g3}). In Fig.3b, the finite value of the coherence
radius was also taken into account. From Eq.(\ref{g3}) one can see
that the maximum visibility is achieved when two of the three phases
are varied simultaneously in the opposite directions,
$\phi_{12}=-\phi_{32}$. In accordance with this, in our experiment
the glass plates in front of detectors 1 and 3 were rotated
synchronously, both clockwise (since detector 1 was in the reflected
beam and detector 3 in the transmitted beam, this led to the
opposite variation of the phases). The obtained visibility for the
case of coherent radiation is $74$\%; for the case of pseudo-thermal
radiation, $38$\%. In each plot, we also show the spatial dependence
of single counts for one of the detectors whose phase was scanned.
Although the dependence is not completely flat (the variation is
caused by the speckle structure of the laser light and the envelope
of the single-slit diffraction pattern), the two-slit interference
pattern in the intensity distribution is erased. Note that the
presented ICF was normalized to the product of the three
intensities; as a result, the `noisy' structure of single-photon
counts did not influence the third-order interference pattern.

If the third-order interference pattern is scanned by only one of
the three detectors, the visibility is lower than $81.9$\% but still
considerably higher than in the case of the second-order
interference. It reaches its maximal value $\frac{\sqrt{2}}{2}$
(approximately $70.5$\%) when the relative phase between the
remaining two detectors is equal to $\pi/2$.

Quite similarly, one can show that the fourth-order intensity
interference of two classical sources is given by the formula

\begin{eqnarray}
g_{1234}^{(4)}=\frac{g^{(4)}}{8}+
\frac{g^{(3)}}{2}+\frac{3[g^{(2)}]^2}{8}+
\frac{g^{(3)}+[g^{(2)}]^2}{4}[\cos\phi_{12}\nonumber
\\
+\cos\phi_{13}+\cos\phi_{14}+
\cos(\phi_{12}-\phi_{13})+\cos(\phi_{12}-\phi_{14})\nonumber
\\
+ \cos(\phi_{13}-\phi_{14})]+
\frac{[g^{(2)}]^2}{8}[\cos(\phi_{12}+\phi_{13}-\phi_{14})\nonumber
\\
+\cos(\phi_{12}+\phi_{14}-\phi_{13})+\cos(\phi_{13}+\phi_{14}-\phi_{12})].
\label{g4}
\end{eqnarray}
Here, the same notation for the phases and normalized ICFs is used
as in Eq.(\ref{g3}). Analysis of this expression shows that the
maximal visibility of the fourth-order interference for thermal
sources is $77.8$\% while for coherent sources, it is $94.4$\%. The
last figure exceeds the visibility values achieved in nearly all
known `four-photon' experiments.

Experimental observation of the fourth-order interference by
registering four-fold coincidences was difficult because of the low
coincidence counting rate in this case. For this reason, we turned
to another method of measuring spatial ICFs, the one based on
digital image processing (see, for instance,
Ref.~\cite{Lugiato_CCD}). The interference pattern in the far-field
zone was registered by a digital photographic camera Canon PowerShot
S2 IS. The source, again a frequency-doubled YAG:Nd laser, in this
case had a repetition rate $50$ Hz. Each frame was made with a
single laser pulse, the exposure time (1/60 s) being less than the
distance between the pulses. A typical interference pattern recorded
in one frame is shown in Fig.4. To accumulate sufficient statistics,
$n=500$ shots were made. Due to the phase shift introduced by the
EOM, the phase of the interference pattern varied from frame to
frame, so that the intensity spatial distribution averaged over all
frames revealed almost no interference (the visibility of Young's
interference pattern was less than $10$\%). In each frame, a
rectangular area was selected, with the dimensions $50$ pixels in
the vertical ($y$) direction and $600$ pixels in the horizontal
($x$) direction (a rectangle shown in Fig.4). First, the intensity
distribution recorded in each frame was averaged over the $y$ side
of the rectangle, so that intensity distributions $I_j(x)$ for all
$n$ pulses were obtained. Next, averaging over $j$ was performed,
the averaged intensity distribution and the third- and fourth-order
normalized correlation functions being calculated as

\begin{equation}
I(x)=\langle I_j(x)\rangle\equiv\frac{1}{n}\sum_{j=1}^n I_j(x),
\label{calc1}
\end{equation}

\begin{equation}
g^{(3)}(x)=\frac{\langle
I_j(x)I_j(0)I_j(-x)\rangle}{I(x)I(0)I(-x)},\label{calc2}
\end{equation}

\begin{equation}
g^{(4)}(x)=\frac{\langle
I_j(x)I_j(0)I_j(-x)I_j(-2x)\rangle}{I(x)I(0)I(-x)I(-2x)}.
\label{calc3}
\end{equation}

Note that the arguments of intensities used in (\ref{calc2}),
(\ref{calc3}) are chosen so as to provide the maximal visibility. In
the above-described coincidence method of $g^{(3)}$ measurement,
this was achieved by scanning two detectors in opposite directions
and the third one being fixed; for a $g^{(4)}$ measurement, the
fourth detector should be scanned with a double speed.

Fig.5 shows the third-order (a) and fourth-order (b) interference
patterns obtained for coherent sources by means of digital-image
processing. For convenience and similarity with Fig.3, the $x$
coordinate, originally measured in pixels, is plotted in phase
units, so that $x=\phi_{12}$; the other phases obey the relations
$\phi_{32}=-\phi_{12}, \phi_{42}=-2\phi_{12}$. As expected, the
distribution in Fig 5a is similar to the third-order interference
pattern registered by means of coincidence method. The interference
visibilities achieved for three- and four-photon interference are
$73\%$ and $93\%$, respectively. For comparison, theoretical
dependencies given by Eqs (\ref{g3}), (\ref{g4}) are shown as thin
solid lines. The averaged distribution of the intensity (shown in
both plots by empty circles) has almost no modulation with the
period of the interference pattern; the observed non-uniformity is
caused by the interference pattern envelope and the edge effects in
the EOM.

Similar dependencies (with smaller visibilities) were measured for
the case of pseudo-thermal sources.

In conclusion, we have demonstrated, both theoretically and
experimentally, that while the classical visibility limit for
two-photon interference is only $50$\%, it becomes considerably
higher in the multi-photon case. Three-photon and four-photon
interference has been observed both for coherent and pseudo-thermal
light, and the maximal visibility values $74$\% and $93$\% have been
achieved. These results, on the one hand, show that classical pulsed
radiation, in addition to the high spatial resolution of ghost
diffraction and ghost imaging, can also provide a high visibility,
but in this case multi-photon correlations should be used. On the
other hand, our results establish a  visibility `threshold' in
multi-photon interference experiments: for the interference to be
clearly nonclassical, the visibility should exceed $81.9$\% in the
three-photon case and $94.4$\% in the four-photon one. Since
exceeding these rather high values is practically difficult, we
suggest that the visibility criterion should be simply avoided in
multi-photon interference experiments, and some other observable
signs of nonclassicality be used, such as the Lee-Klyshko
criterion~\cite{Lee-klyshko}, relations between the different-order
normalized ICFs~\cite{NL}, or the scaling of normalized ICFs with
the mean photon number~\cite{quantel}. This work was supported in
part by the RFBR grants \# 06-02-16393, \# 05-02-16391, \#
06-02-39015-GFEN and the Program of Leading Scientific Schools
Support, \# NSh-4586.2006.2. T.Sh.I. acknowledges the support of the
'Dynasty' Foundation.

\begin{figure}
\includegraphics[height=2cm]{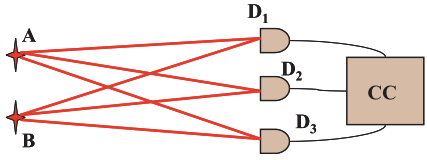} \caption{Observation of three-photon interference in
the Young double-slit scheme.}
\end{figure}

\begin{figure}
\includegraphics[height=4cm]{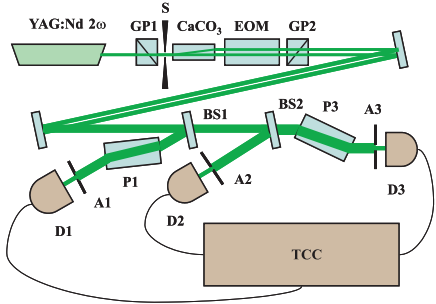}
\caption{Experimental setup. GP1 and GP2, Glan prisms; S, a slit;
EOM, electro-optic modulator; BS1, 33\%/67\% beamsplitter; BS2, 50\%
beamsplitter; A1-3, apertures; P1,3, glass plates; D1-3, avalanche
photodiodes; TCC, triple coincidence circuit.}
\end{figure}

\begin{figure}
\includegraphics[height=4cm]{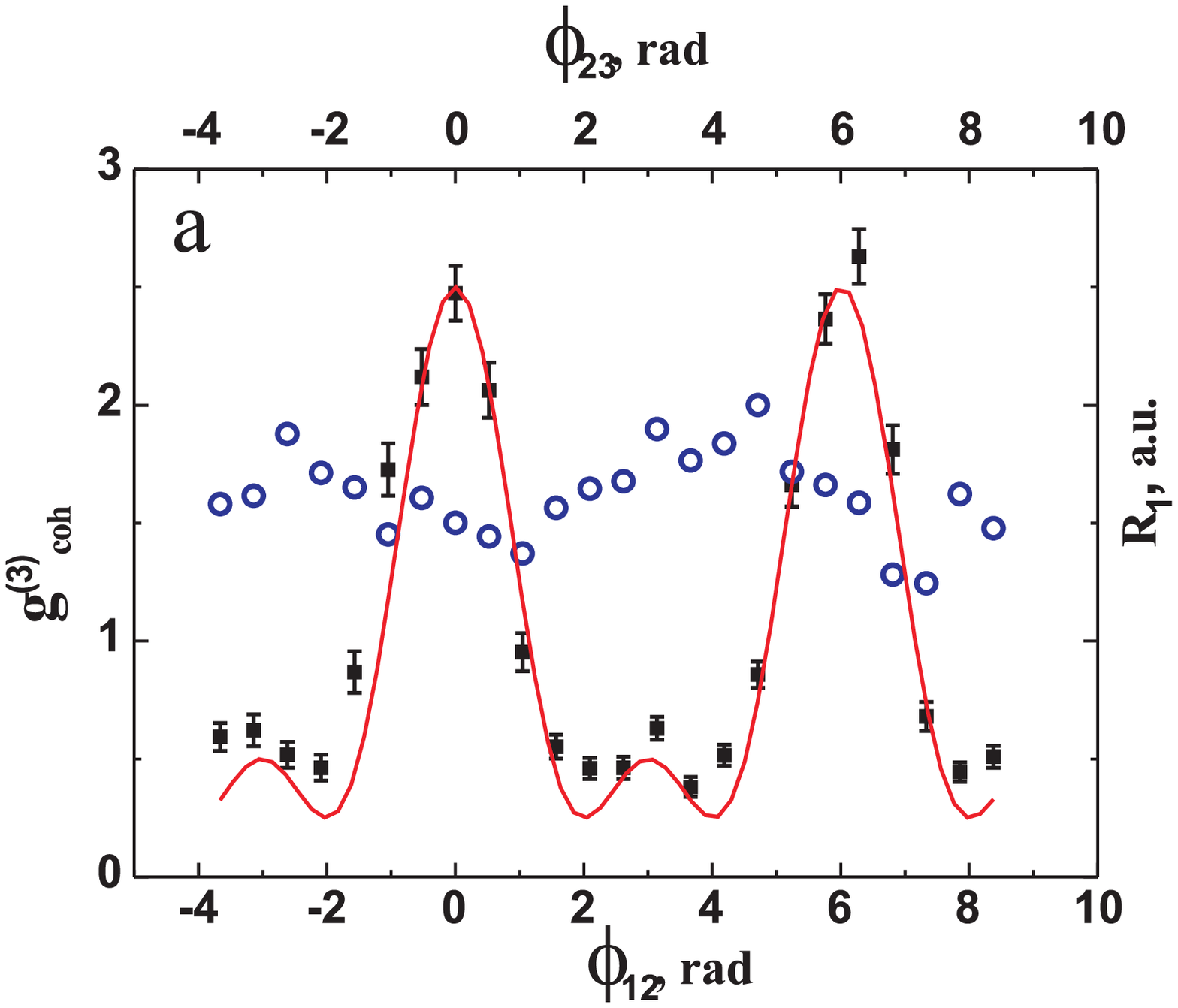}
\includegraphics[height=4cm]{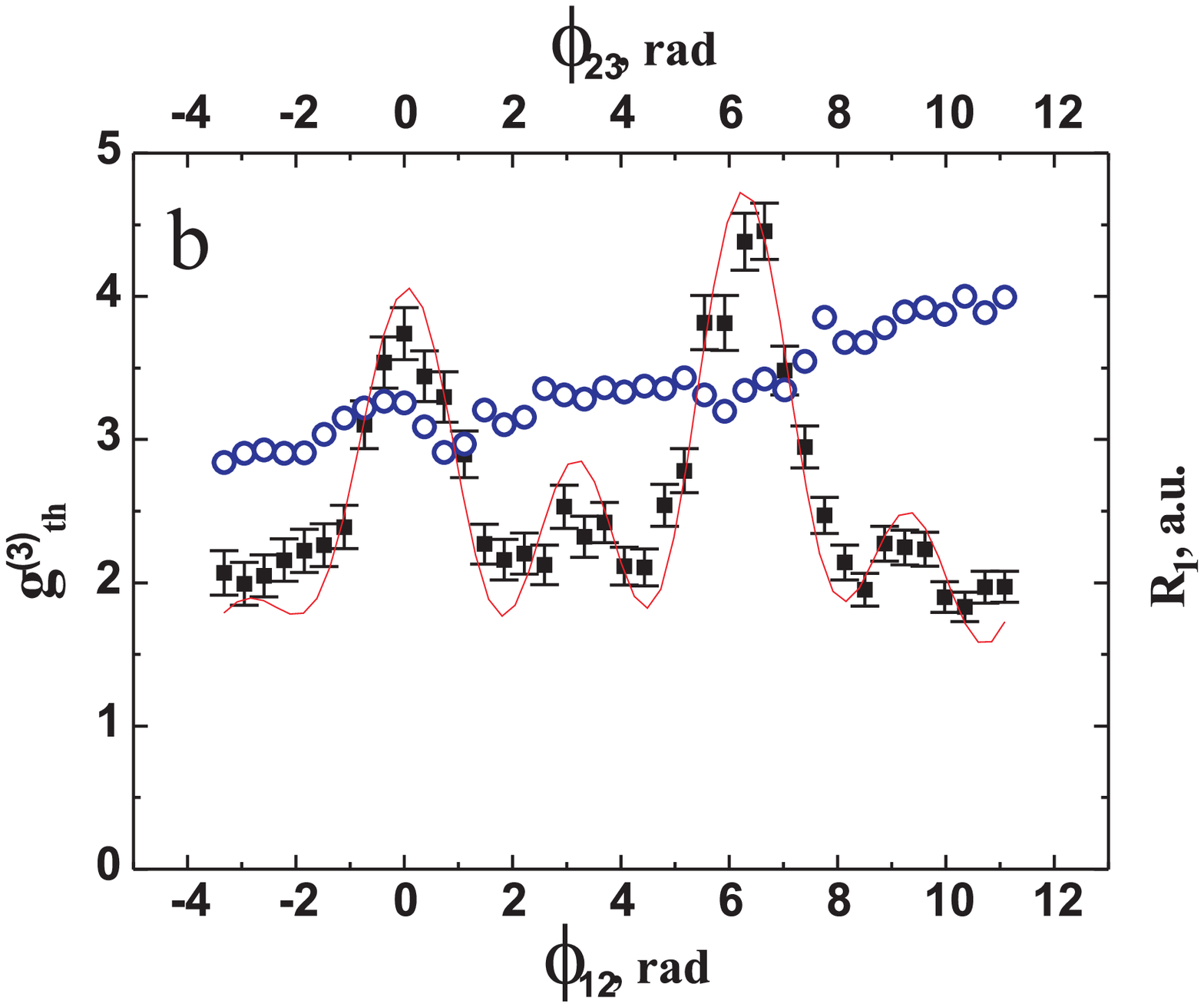}
\caption{Interference pattern in the normalized third-order ICF for
(a) coherent sources (b) pseudo-thermal sources obtained by tilting
glass plates at the inputs of detectors 1 and 3. Empty circles show
the intensity distribution given by the counting rate $R_1$ of
detector 1. Solid lines show the theoretical fit with Eq.(\ref{g3})}
\end{figure}

\begin{figure}
\includegraphics[height=3cm]{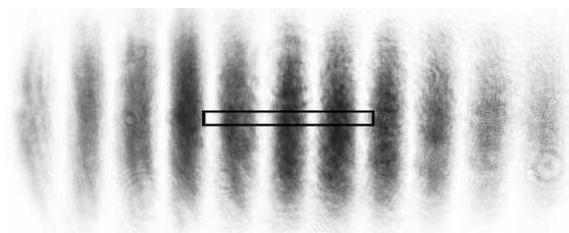}
\caption{Single-pulse photograph of the Young interference pattern
for two coherent sources. The rectangle shows the dimensions of the
processed area.}
\end{figure}

\begin{figure}
\includegraphics[height=3cm]{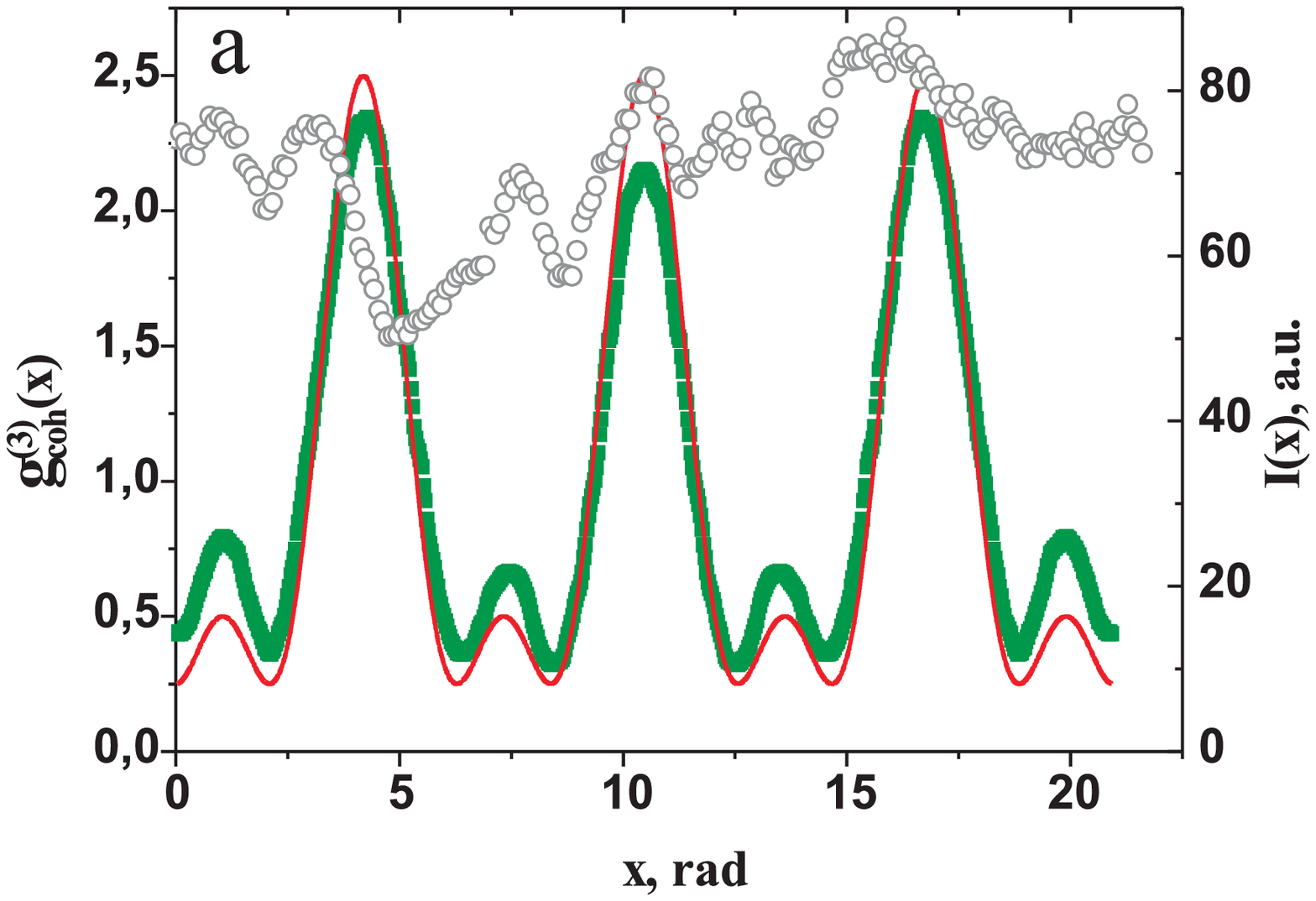}
\includegraphics[height=3cm]{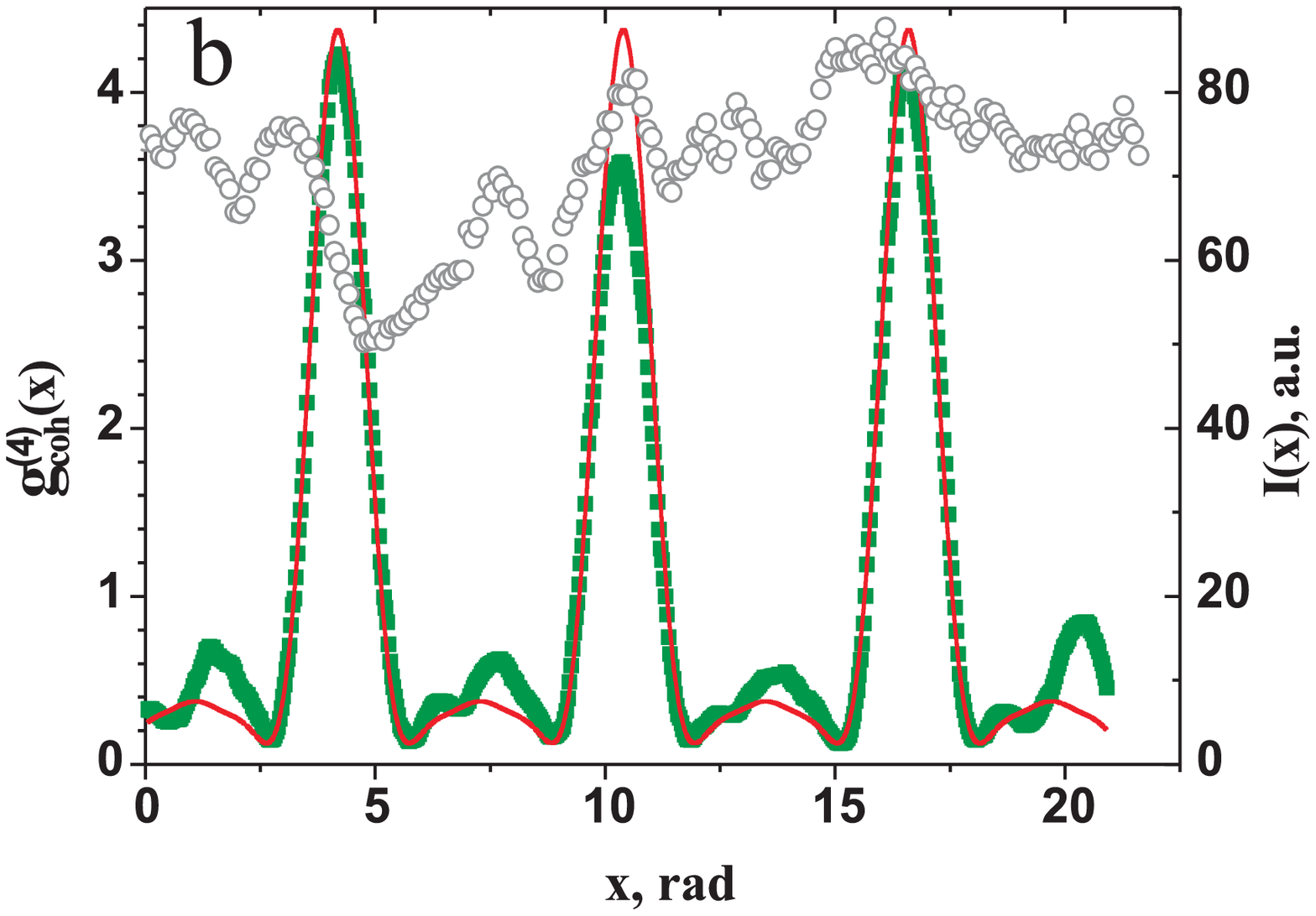}
\caption{Interference patterns in  normalized third-order (a) and
fourth-order (b) ICFs for coherent sources obtained by processing
digital images as shown in Fig.4. The averaged intensity
distribution in the $x$ coordinate is shown in both plots by empty
circles. Thin solid lines show the theoretical dependencies
(\ref{g3}) (Fig.5a) and (\ref{g4}) (Fig.5b).}
\end{figure}


\begin{references}
\bibitem{generality}Description of other types of two-beam
interferometers such as, for instance, Mach-Zehnder one, is similar.
\end{references}
\end{document}